\documentclass[aps,prd,email,preprint,showpacs,showkeys,preprintnumbers,amsmath,amssymb,nofootinbib,superscriptaddress]{revtex4}

\usepackage[dvips]{graphicx}
\usepackage{color}
\usepackage{latexsym}
\usepackage{amsmath}
\usepackage{amssymb}
\usepackage{eufrak}
\usepackage{euscript}
\usepackage{pstricks}
\usepackage{graphics}
\usepackage{graphicx}
\usepackage{picture}

\newcommand{\be}{\begin{equation}}
\newcommand{\ee}{\end{equation}}
\newcommand{\bn}{\begin{eqnarray}}
\newcommand{\en}{\end{eqnarray}}

\newcommand{\p}{\partial}

\newcommand{\de}{\delta}

\def\no{\nonumber}

\def\[{\left\lbrack}
\def\]{\right\rbrack}

\def\({\left(}
\def\){\right)}
\def\ni{\noindent}	
\def\MyItem[#1]#2{\item[{#1}]#2}

\def\[{\left\lbrack}
\def\]{\right\rbrack}

\def\({\left(}
\def\){\right)}

\begin{document}
%%%%%%%%%%%%%%%%%%%%%%%%%%%%%%%%%%%%%%%%%%%%%%%%%%%%%%%%%%%%%%%%%%%%%%%%%%%%%%%%%%%%%%%%%%%
%%%%%%%%%%%%%%%%%%%%%%%%%%%%%%%%%%%%%%%%%%%%%%%%%%%%%%%%%%%%%%%%%%%%%%%%%%%%%%%%%%%%%%%%%%%%%%
%%%%%%%%%%%%%%%%%%%%%%%%%%%%%%%%%%%%%%%%%%%%%%%%%%%%%%%%%%%%%%%%%%%%%%%%%%%%%%%%%%%%%%%%%%%%%%
%\begin{center}

%\Large{\bf Some considerations about Jackiw-Pi model}
%\Large {\bf New aspects of quantization of Jackiw-Pi model: field-antifield formalism and noncommutativity}

%\bigskip 
%\bigskip

%{\small
%Vahid Nikoofard$^{a,c}$\footnote{vahid@cbpf.br} and
%Everton M. C. Abreu$^{b,c}$\footnote{evertonabreu@ufrrj.br}\\ \bigskip 

%{\small
%$^{a}$LAFEX, Centro Brasileiro de Pesquisas F\'{i}sicas, Rua Xavier Sigaud 150, 22290-180, \mbox{Rio de Janeiro}, RJ, Brazil\\
%$^{b}$Grupo de F\'{i}sica Te\'orica e Matem\'atica F\'{i}sica, Departamento de F\'{i}sica, Universidade Federal Rural do Rio de Janeiro, 23890-971, Serop\'edica, %RJ, Brazil\\
%$^{c}$Departamento de F\'{i}sica, Universidade Federal de Juiz de Fora, 36036-330, Juiz de Fora, MG, Brazil\\}
%\end{center}

%\date{\today}
%%%%%%%%%%%%%%%%%%%%%%%%%%%%%%%%%%%%%%%%%%%%%%%%%%%%%%%%%%%%%%%%%%%%%%%%%%%%%%%%%%%%%%%%%%%%%%%%%%%%%%%%%%%%%%%%%%%%
%%%%%%%%%%%%%%%%%%%%%%%%%%%%%%%%%%%%%%%%%%%%%%%%%%%%%%%%%%%%%%%%%%%%%%%%%%%%%%%%%%%%%%%%%%%%%%%%%%%%%%%%%%%%%%%
%%%%%%%%%%%%%%%%%%%%%%%%%%%%%%%%%%%%%%%%%%%%%%%%%%%%%%%%%%%%%%%%%%%%%%%%%%%%%%%%%%%%%%%%%%%%%%%%%%%%%%%%%%%%%%%%
\title{\Large{New aspects of quantization of Jackiw-Pi model: field-antifield formalism and noncommutativity}}

\author{Vahid Nikoofard}\email{vahid@cbpf.br}
\affiliation{LAFEX, Centro Brasileiro de Pesquisas F\'{i}sicas, Rua Xavier Sigaud 150, 22290-180, \mbox{Rio de Janeiro}, RJ, Brazil}
\author{Everton M. C. Abreu} \email{evertonabreu@ufrrj.br}
\affiliation{Grupo de F\' isica Te\'orica e Matem\'atica F\' isica, Departamento de F\'{\i}sica,
Universidade Federal Rural do Rio de Janeiro,
BR 465-07, 23890-971, Serop\'edica, RJ, Brazil}
\affiliation{Departamento de F\'{\i}sica, ICE, Universidade Federal de Juiz de Fora,
36036-330, Juiz de Fora, MG, Brazil}

\date{\today}

%%%%%%%%%%%%%%%%%%%%%%%%%%%%%%%%%%%%%%%%%%%%%%%%%%%%%%%%%%%%%%%%%%%%%%%%%%%%%%%%%%%%%%%%%%%%%%%%%%%%%%%%%%%
%%%%%%%%%%%%%%%%%%%%%%%%%%%%%%%%%%%%%%%%%%%%%%%%%%%%%%%%%%%%%%%%%%%%%%%%%%%%%%%%%%%%%%%%%%%%%%%%%%%%%%%%%%%

\begin{abstract}
\ni The so-called Jackiw-Pi (JP) model for massive vector fields is a three dimensional, gauge invariant and parity preserving model which was discussed in several contexts.  In this paper we have discussed its quantum aspects through the introduction of Planck scale objects, i.e., via noncommutativity and the well known BV quantization.  Namely, we have constructed the JP noncommutative space-time version and we have provided the BV quantization of the commutative JP model and we have discussed its features.  The noncommutativity has introduced interesting new objects in JP's Planck scale framework.  
\end{abstract}


\maketitle
%\tableofcontents

\section{Introduction}

It is well known that quantum field theories are plagued with infinities that turns the calculations of some physical objects simply impossible.  Some techniques in order to tame such infinities like renormalization have their effectiveness limited and other alternatives such as supersymmetry are still under investigation and needs some experimental observations in order to lead the experts to the undoubtedly direction.  However, the idea that the change of a continuum spacetime to a discrete one has bring some light that has been explored since the recent results connecting noncommutative (NC) geometry to string theory embedded within a magnetic background \cite{sw}.  However, this fuzzy spacetime concept was firstly published by Snyder a long time ago \cite{snyder} in order to free QFT from the infinities, but unfortunately, it did not succeed \cite{yang}.  After its connection to string theory, a lot of interest is increasing and interesting results were obtained \cite{reviews1,reviews2}.

However, some problems concerning unitarity  and causality \cite{unitarity} of NC theories are still present but it is possible to live with them.  For example, soliton solutions in NC systems have originated interesting analysis since we can have an emergence of a length scale induced by the NC parameter ($\theta$) are non-local and this has generated solitons in NC field theories even though in some cases solitons are absent in the respective
theories in ordinary spacetime. We can cite as an example the NC solitons in higher dimensional scalar field theories \cite{gms}.  The investigation of NC soliton and magnetic coupling can be found in \cite{ghosh}.

One of the reasons that makes three dimensional gauge theories still have theoretical (mathematical) interest is because they describe (1) kinematic processes that are confined to a plane when external structures (magnetic fields, cosmic strings) perpendicular to the plane are present, and (2) static properties of (3 + 1)-dimensional systems in equilibrium with a high temperature heat bath. In condensed matter physics, they describe the topological order in fractional quantum Hall effect states. An important issue is whether the apparently massless gauge theory possesses a mass gap. The suggestion that indeed it does gain support from the observation that the gauge coupling constant squared carries dimension of mass, thereby providing a natural mass-scale (as in the two-dimensional Schwinger model) \cite{Jackiw:1997ha}.
Also, without a mass gap, the perturbative expansion is infrared divergent, so if the theory is to have a perturbative definition, infrared divergences must be screened, thereby providing evidence for magnetic screening
in the four-dimensional gauge theory at high temperature.

The Chern-Simons term as a topological theory of Schwarz type when added to the three-dimensional Yang-Mills action, renders the fields massive, while preserving gauge invariance. However, the drawback of the Chern-Simons topological mass term is the loss of parity-invariance, due to the presence of the $\epsilon^{\mu\nu\rho}$-tensor. A trivial
way of maintaining parity with this mass generation is through the doublet mechanism. Consider a pair of identical Yang-Mills actions, each supplemented with their own Chern-Simons term, which enters with opposite signs. The parity transformation is defined to include field exchange accompanying coordinate reflection, and this is a symmetry of the doubled theory. Using this method Jackiw and Pi in a seminal paper \cite{Jackiw:1997jga} have offered a theory for massive vector fields, which is gauge invariant and parity preserving. This theory is gauge invariant, but has non-Yang-Mills dynamics. Although formal quantization of the model can be carried out, developing a perturbative computation method encounters some difficulties.

The consistency of physical states of different types of Jackiw - Pi model was studied using constrained analysis in the Hamiltonian approach \cite{Dayi:1997in}, and classical symmetries using the algebraic non-perturbative method were discovered in the BRST formulation \cite{DelCima:2011bx}. Based on the Bonora-Tonin superfield formalism \cite{Bonora:1980pt}, (anti)BRST-symmetry of JP-model was analyzed in \cite{Gupta:2011cta}. The classical characteristics of the model, such as BRST invariance, gauge-fixing, and Slavnov-Taylor identity were studied in \cite{DelCima:2012bm}. In 3D Schouten-ghost-free gravity, in the Hamiltonian formalism, Deser, Ertl and Grumiller \cite{Deser:2012ci} have demonstrated the bifurcation effect, namely, the clash between two local invariances. It is conjectured that such a bifurcation effect could appear in the JP model, since it conforms two local invariances. The importance of JP-model also can be found in a different context. It has been conjectured that the superalgebra OSp$(32\lvert1)$ is the full symmetry group of M-theory \cite{Hull:1994ys}. It was pointed out in \cite{Smolin:2000kc} that Chern - Simons theory for the superalgebra $OSp(32\lvert1)$ contains the so-called M-theory matrix models. Therefore the aforementioned advantage of JP-model over Chern-Simons theory mandates the supersymmetrization of the original JP-model \cite{Nishino:2015hha}. An extension of the JP model with the inclusion of a new kinetic term was studies in \cite{Nishino:2015hha}.

As the JP model is non-Abelian, we can not construct its noncommutative (NC) counterpart by simply substituting the dot product by the star one and using the Seiberg-Witten  (SW) map. Generally in the common method one assumes $U(1)$ as the gauge group \cite{Banerjee:2005zq}. Although it must be mentioned that $U(N)$ is a non-Abelian group but we can analyze it by the standard method. But for an arbitrary gauge group the
commutation of two gauge transformations is not another gauge transformation of the same group \cite{Jurco:2001rq}. It will be closed in only the enveloping algebra of the original algebra.\\
Here we try to construct the NC counterpart of the model proposed by Jackiw and Pi for an arbitrary gauge group using the enveloping algebra of the original algebra. For this reason we have used a method elaborated by J. Wess et al. \cite{Jurco:2001rq}. The generalization of this method to higher order term of NC parameter can be found in a work done by Ukler et al. \cite{Ulker:2007fm}. In this work we have just proceeded up to the first order term in our calculations. 

The Batalin-Vilkovisky (BV) or field-antifield formalism \cite{Batalin:1981jr} is until now the most complete method to deal with quantum gauge field theory. In fact it is a generalization of the BRST formalism \cite{Becchi:1974xu,Tyutin1975}
that includes the anti-fields sources into the action. One of the reasons theoretical physicists are interested in a BRST invariant action is that it leads to Slavnov-Taylor identities from which one may prove unitarity and renormalizability. Among the various
BRST approaches, the BV formalism has the advantage of treating all quantum systems (with/without open algebra\textquoteright{}s, with/without ghosts for ghosts) in a unified manner. This brings out the essential
features more clearly, and that, in turn, might be helpful in quantizing systems, such as the heterotic string or closed-string field theory. In some sense, the BV formalism is a generalization of BRST quantization. In fact, when sources of the BRST transformations are introduced into the configuration space, the BRST approach resembles the field-antifield one \cite{Weinberg:1996kr}. Antifields then, have a simple interpretation: they are the sources for BRST transformations. In this sense, the field-antifield formalism is a general method for dealing with gauge theories within the context
of standard field theory.

The general structure of the antibracket formalism is as follows. One introduces an antifield for each field and ghost, thereby doubling the total number of original fields. The antibracket \mbox{( , )} is an odd non-degenerate symplectic form on the space of fields and antifields. The original classical action $S_{0}$ is extended to a new action $S$, in an essentially unique way, to arrive at a theory with manifest BRST symmetry. One equation, the master equation $\left(S,S\right)=0$, reproduces in a compact way the gauge structure of the original theory governed by $S_{0}$. Although the master equation resembles the Zinn-Justin
equation, the content of both is different since $S$ is a functional of quantum fields and antifields and is a functional of classical fields.

In this work after studying carefully the gauge structure of Jackiw-Pi (JP) model, we will construct the corresponding BV action for the $U(1)*U(1)*U(1)*SU(N)$ gauge group. It is obvious that the quantization of this gauge group is possible via BRST approach but we have hired the BV formalism for having better understanding of its symmetries. Also gauge fixing is simpler in this formalism and moreover the BV action is ready for quantization and study of anomalies. 

The issues dealt in this paper follows the sequence such that in section 2 we have discussed the JP model and its NC version was constructed in section 3.  Concerning quantization, we have carried out the field-antifield of the extended JP model in section 4.  The conclusions are depicted in section 5.  In order to try to keep the paper self-contained, we have added two Appendices with brief reviews of the SW mapping and the basics of the BV quantization. 

%%%%%%%%%%%%%%%%%%%%%%%%%%%%%%%%%%%%%%%%%%%%%%%%%%%%%%%
%%%%%%%%%%%%%%%%%%%%%%%%%%%%%%%%%%%%%%%%%%%%%%%%%%%%%%%%%%%

\section{The Jackiw-Pi Theory}

The JP model is a non-Abelian gauge invariant, massive, parity preserving theory governed by the Lagrangian \cite{Jackiw:1997ha, Jackiw:1997jga}
\be 
\label{jp}
\mathcal{S}=Tr\int d^{3}x\left(\frac{1}{2}F^{\mu\nu}F_{\mu\nu}+\frac{1}{2}G^{\mu\nu}G_{\mu\nu}-m\epsilon^{\mu\nu\rho}F_{\mu\nu}\phi_{\rho}\right)
\ee

\ni where $A_{\mu}$ and $\phi_{\mu}$ are vector bosonic fields and $m$ is a mass parameter. The 2-form curvature $F^{\(2\)} = dA^{\(1\)}-i\(A^{\(1\)} \wedge A^{\(1\)}\) =\frac{1}{2!}\(dx^\mu \wedge dx^\nu\) F_{a\mu\nu}\,T^a$ defines
the curvature tensor $F_{\mu\nu} = \p_{[\mu}A_{\nu]} -i\[A_\mu, A_\nu\]$ for the non-Abelian 1-form since $A^{\(1\)}=dx_\mu A_a^\mu \, T^a$.   
The non-Abelian  gauge field $A_\mu=A_{a\mu}T^a$ where $d=dx^\mu\p_\mu$ is the exterior derivative (with $d^2=0$). Similarly, another 2-form $G^{\(2\)} = d\phi^{\(1\)} -i\(A^{\(1\)} \wedge \phi^{\(1\)}\) -i\(\phi^{\(1\)} \wedge A^{\(1\)}  \)  =\frac{1}{2!}\(dx^\mu \wedge dx^\nu\) G_{a\mu\nu} \, T^a$ defines the curvature tensor 
\bn
\label{AA}
G_{\mu\nu}^{ab}=D_{\mu}^{ab}\phi_{\nu}-D_{\nu}^{ab}\phi_{\mu}\,\,,
\en

\ni where the covariant derivative is written in terms of the structure constants $f^{abc}$ as
\bn
\label{A}
D^{ab}_\mu = \delta^{ab} \p_\mu \,+\,\lambda f^{abc} A^c_\mu \,\,.
\en

In Eq. (\ref{AA}) we have the 1-form $\phi^1=dx^\mu \phi_{\mu}$ vector field $\phi_\mu = \phi_{a\mu} T^a$.  In the above, the vector fields $A_\mu$ and $\phi_\mu$ have opposite parity, thus the JP model becomes parity invariant. In this classical theory in commutative spacetime, the fields are Lie algebra-valued $\Psi=\Psi^{a}T^{a}$ but in the NC spacetime for an arbitrary gauge group, as it was mentioned before, this property will be lost.

This theory is invariant under the non-Abelian transformation
\begin{eqnarray} \label{ym-jp}
\delta_{\theta}A_{\mu} &=& D_{\mu}\theta \no \\
\delta_{\theta}\phi_{\mu}&=&-i[\phi_{\mu},\theta].
\end{eqnarray}

\ni and this last mixing term in ${\cal S}$ is also invariant under 
\bn
\label{B}
\delta_2 A_\mu =0 \qquad
\mbox{and}  \qquad
\delta_2 \phi_\mu = D_\mu \chi \,\,.
\en

\ni which brings an interesting problem since this second transformation does not male the non-linear part of $G^{mu\nu}$ invariant, since 
$\delta_2 G^{\mu\nu} = [F^{\mu\nu},\,\chi]$.   Hence, the quadratic theory has two independent and Abelian gauge transformations.  When we consider interaction, one non-Abelian symmetry survives which affects the quantization \cite{Jackiw:1997jga}.  We will come back to this issue in a minute.

The Lie algebra of the generators for the symmetry group of $A_\mu$ is given by 
\be
\label{4}
\left[T^{a},T^{b}\right]=if^{abc}T^{c}
\ee
and we recall that the vector potential $A_{a\mu}$ is the connection associated with this group. The gauge group of $\phi_{\mu}$ is Abelian and its generators are symmetric matrices with the same number of generators as $A_{\mu}$ and they obey the following commutation relationship
\be
\label{5}
\left[P^{a},P^{b}\right]=0.
\ee
Also, it is assumed that the generators of these two algebra satisfy the following relation
\be
\label{6}
\left[T^{a},P^{b}\right]=if^{abc}P^{c}.
\ee

\ni In the case of $su(n)$, the generators of the Lie algebra are traceless and Hermitian matrices.  Also we will assume that the generators $P^a$ are symmetric matrices.

By turning off the coupling to $\lambda$ in Eq. (\ref{A}), we have that 
\bn 
\label{abelian_jp}
\mathcal{S}_q \equiv \mathcal{S}\quad  (\lambda=0)
\en

\ni and the action in Eq. (\ref{jp}) reduces to an action which is invariant under two different Abelian transformations
\bn 
\label{abelian_tr_jp}
\delta_{q1}A_\mu &=& \p_\mu \theta \qquad ; \qquad \delta_{q1}\phi_\mu = 0 \no \\
\delta_{q2}A_\mu &=& 0 \qquad ; \qquad  \delta_{q2}\phi_\mu =\p_\mu \xi. 
\en 

\ni For the Green functions generating functional (or the partition function) we just need the gauge fixing terms for its gauge symmetries of the Eq. (\ref{ym-jp}). However, the propagators will be calculated in terms of a quadratic action Eq. (\ref{abelian_jp}) which still possesses the gauge symmetry of the Eq. (\ref{abelian_tr_jp}), i.e., the gauge fixing of the non-Abelian action will not be enough to eliminate the superficial fields in Eq. (\ref{abelian_jp}) which are essential to define finite propagators.

A general quantization procedure of the theories whose gauge symmetries in the quadratic (${\cal S}_q$) and the full (non-Abelian) cases are not consistent, is not available yet \cite{Dayi:1997in}. Jackiw and Pi proposed to enlarge the configuration space by introducing the new
fields $\rho$ and to deal with the action (Extended JP model)
\be
\label{extended}
\mathcal{S}_{ext}=Tr\int d^{3}x\left(\frac{1}{2}F^{\mu\nu}F_{\mu\nu}+\frac{1}{2}\left(G^{\mu\nu}-i\left[F^{\mu\nu},\rho\right]\right)\left(G_{\mu\nu}-i\left[F_{\mu\nu},\rho\right]\right)-m\epsilon^{\mu\nu\lambda}F_{\mu\nu}\phi_{\lambda}\right)
\ee

\ni where $\rho=\rho^a T^a$ and which is invariant under two different type of non-Abelian transformations
\begin{eqnarray} \label{ym-jp2}
\text{Yang-Mills}\left\{ \begin{array}{c}
\delta_{\theta}A_{\mu}=D_{\mu}\theta\\
\delta_{\theta}\phi_{\mu}=-i[\phi_{\mu},\theta]\\
\delta_{\theta}\rho=-i[\rho,\theta]
\end{array}\right.
\end{eqnarray}

\ni and

\begin{eqnarray} \label{non-ym-jp}
\text{Non-Yang-Mills}\left\{ \begin{array}{c}
\delta_{\chi}A_{\mu}=0\\
\delta_{\chi}\phi_{\mu}=D_{\mu}\chi\\
\delta_{\chi}\rho=-\chi
\end{array}\right.
\end{eqnarray} 
The additional scalar field $\rho$ transforms under the first gauge transformation as an adjoint vector while the second one applies a shift.  It is very important to understand that the action in Eq. (\ref{jp}) describes ``charged vector mesons," represented by $\phi_\mu$, and we can see that it interacts minimally with the gauge potential $A_\mu$.   In order to construct an action which incorporates both gauge transformations (\ref{ym-jp}) and (\ref{B}), it is a common practice to introduce (as in (\ref{jp})) additional non-minimal interactions.   This new extended version of Eq. (\ref{jp}), Eq. (\ref{extended}), is invariant under these both gauge transformations which are incorporated in a larger non-Abelian Yang-Mills gauge symmetry \cite{Jackiw:1997jga}.   
Besides, we can combine (\ref{ym-jp}) and (\ref{B}) into a non-Abelian gauge symmetry without changing the dynamics.   Hence, Jackiw and Pi introduced a new scalar field multiplet $\rho$ (Eq.(\ref{extended})).   The new gauge transformations of (\ref{extended}) are given by (\ref{ym-jp2}) and (\ref{non-ym-jp}) and the Abelian and non-Abelian generators are in Eqs. (\ref{4}), (\ref{5}) and (\ref{6}).  Notice the (\ref{extended}) is invariant, although it does not have the Yang-Mills form.  It is always possible to set $\rho=0$ in order to regain the dynamics of (\ref{jp}).  Jackiw and Pi have shown that the quantization is straightforward
\cite{Jackiw:1997jga}.  We will show here that the introduction of a Planck scale parameter is possible, which is equivalent to a semi-classical treatment of the extended as well as the NC version of the $\rho=0$ (non-extended) model.   Since both are connected via $\rho$, this NC result shows that the connection is preserved.  Namely, the NC contribution does not modify the link between (\ref{jp}) and (\ref{abelian_tr_jp}).

In this work we have discussed the NC alternatives of both versions of the JP model, i.e., the non-extended (Eq. (\ref{jp})) and the extended (Eq. (\ref{extended})) one.  However, it is also important to notice that the non-extended one is recovered when we make $\rho=0$ in Eq. (\ref{extended}) as well as we have this equivalence  concerning the symmetries of the extended model in Eqs. (\ref{ym-jp2}) and (\ref{non-ym-jp}). Namely, the symmetries (\ref{ym-jp})-(\ref{B}) can be recovered from (\ref{ym-jp2})-(\ref{non-ym-jp}) when $\rho=0$ in this last set of symmetries.
Although we have introduced noncommutativity in both versions, when we will analyze the BV quantization, due to the analogousness of the computation, we will show the BV quantization of the extended JP model.  The result for the non-extended one can be directly obtained by considering the above $\rho=0$ substitution in the action and symmetries.  We will mention this observation later again.

%%%%%%%%%%%%%%%%%%%%%%%%%%%%%%%%%%%%%%%%%%%%%%%%%%%%%%%%%%%%%%%%%%%%%%%%%%%%%%%%%%%%%%%%%%%%%%%%%%%%%%%
%%%%%%%%%%%%%%%%%%%%%%%%%%%%%%%%%%%%%%%%%%%%%%%%%%%%%%%%%%%%%%%%%%%%%%%%%%%%%%%%%%%%%%%%%%%%%%%%%%%%%%%%%%%
\section{Field-antifield treatment of extended Jackiw-Pi model}

According to the gauge transformations, Eqs. (\ref{ym-jp2}) and (\ref{non-ym-jp}) the gauge structure of extended JP model can be expressed in a compact form $\delta\Psi^{i}=R_{\alpha}^{i}\varepsilon^{\alpha}$ or 

\be
\begin{array}{cccc}
	\left(\begin{array}{c}
		\delta A_{\mu}\\
		\delta\phi_{\mu}\\
		\delta\rho
	\end{array}\right) & = & \left(\begin{array}{cc}
	D_{\mu} & \quad \:0\\
	\left[\phi_{\mu},\bullet\right] & \quad \:D_{\mu}\\
	\left[\rho,\bullet\right] & \quad \:-1
\end{array}\right) & \left(\begin{array}{c}
\theta\\
\chi
\end{array}\right)\end{array}.
\ee
The dynamical variables of the model, i.e., $A_{\mu}$, $\phi_{\mu}$ and $\rho$ are bosonic fields so their Grassmann parity is $\epsilon_i=0$. The gauge parameters $\theta$ and $\chi$ also are bosonic variables hence their Grassmann parity is  $\epsilon_\alpha=0$.

For the first step we have to calculate the commutation of two gauge transformations. For the gauge
field $A_{\mu}$ we have
\be
\left[\delta_{1},\delta_{2}\right]A_{\mu}^{a}=\partial_{\mu}\theta_{12}^{a}+f^{ade}A_{\mu}^{d}\theta_{12}^{e}=D_{\mu}^{ae}\theta_{12}^{e}
\ee
where $\theta_{12}^{e}=f^{ecb}\theta_{1}^{c}\theta_{2}^{b}$. For the vector field $\phi_{\mu}$ one finds
\be
\left[\delta_{1},\delta_{2}\right]\phi_{\mu}^{a}=f^{adb}\phi_{\mu}^{d}\left(f^{bec}\theta_{1}^{e}\theta_{2}^{c}\right)+D_{\mu}^{ad}\left(f^{dbc}\chi_{1}^{b}\theta_{2}^{c}+f^{dcb}\theta_{1}^{c}\chi_{2}^{b}\right)
\ee
Additionally for the scalar field $\rho$, we yield
\be
\left[\delta_{1},\delta_{2}\right]\rho^{a}=f^{adb}\rho^{d}\left(f^{bec}\theta_{1}^{e}\theta_{2}^{c}\right)-\left(f^{abc}\chi_{1}^{b}\theta_{2}^{c}+f^{acb}\theta_{1}^{c}\chi_{2}^{b}\right).
\ee

\ni As we can see from the commutations of fields, the gauge algebra of the extended JP model is closed and all of $E_{\alpha\beta}^{ij}$ (see Eq. (B5) in Appendix B) are equal to zero. In other words, there is not any term dependent on the equation of motion. The next step would be to determine the structure constants of the
gauge algebra according to Eq. (\ref{eq:commug}). As an interesting result we find that the non-zero structure constant of all the above commutations are the same and are equal to $T_{\beta\gamma}^{\alpha}=f^{abc}$ (Eq. (B5)). 

Now we have the enough ingredients to construct the field-antifield action for the theory at hand as
\begin{eqnarray}
S_{BV} & = & S_{0}+A_{\mu}^{\star a}D^{\mu ab}\xi^{b}+\phi_{\mu}^{\star a}\left(f^{abc}\phi^{\mu b}\xi^{c}+D^{\mu ab}\eta^{b}\right)+\rho^{\star a}\left(f^{abc}\rho^{b}\xi^{c}-\eta^{a}\right)\\
& + & \eta^{\star a}f^{abc}\xi^{c}\xi^{b}+\xi^{\star a}f^{abc}\eta^{c}\xi^{b}
\end{eqnarray}
where $\xi$ and $\eta$ are ghost fields related to the gauge parameters $\theta$ and
$\chi$, respectively. The Grassmann parity of these ghosts is $\epsilon\(\xi\)=\epsilon\(\eta\)=1$. The ghost numbers of the variables of action $S_{BV}$ are
\begin{align}
&\textbf{gh}\[A_\mu\] =\textbf{gh}\[\phi_\mu\]=\textbf{gh}\[\rho\]=0, &&\textbf{gh}\[\xi\] =\textbf{gh}\[\eta\]=1, \no \\
%&\textbf{gh}\[\bar{\xi}\] =\textbf{gh}\[\bar{\eta}\]=-1 &&\textbf{gh}\[\bar{\xi}^\star\] =\textbf{gh}\[\bar{\eta}^\star\]=0,  \\
&\textbf{gh}\[A_\mu^\star\] =\textbf{gh}\[\phi_\mu^\star\]=\textbf{gh}\[\rho^\star\]=-1, &&\textbf{gh}\[\xi^\star\] =\textbf{gh}\[\eta^\star\]=-2.
\end{align}

\ni Before quantization we have to fix the gauge degrees of
freedom. To realize this we go to a gauge-fixed basis by introducing a fermionic function with the ghost number equal to $\textbf{gh}\left[\Theta\right]=-1$ and Grassmann parity $\epsilon\left(\Theta\right)=-1$, as mentioned before. Without lost of generality we suggest the fermionic function
\begin{equation}
\Theta=\int d^3x \: \bar{\xi}^{a}\(-\frac{\bar{\pi}^a}{2\gamma}+\partial^{\mu}A_{\mu}^a\) +\bar{\eta}^{a}\(-\frac{\bar{\omega}^a}{2\gamma^\prime}+\partial^{\mu}\phi_{\mu}^a\)\label{eq:gf}
\end{equation}

\ni where $\bar{\xi}^{a}$ and $\bar{\eta}^{a}$ are Faddeev-Popov
antighost fields related to the ghosts $\xi^{a}$ and $\eta^{a}$
with statistics and ghost number equal to
\be
\epsilon\left(\bar{\xi}^{a}\right)=\epsilon\left(\bar{\eta}^{a}\right)= 1,\hspace{1.5cm}  \textbf{gh}\left[\bar{\xi}^{a}\right]=\textbf{gh}\left[\bar{\eta}^{a}\right]=-1.
\ee
It should be mentioned that the final result of a quantization process is independent
of gauge fixing. Together with the Faddeev-Popov antighost,
we have introduced the Nakanishi-Lautrup fields $\left(\bar{\pi}^{a},\bar{\omega}^{a}\right)$ to
our minimal set to eliminate antighost fields with the following properties
\be
\epsilon\left(\bar{\omega}^{a}\right)=\epsilon\left(\bar{\pi}^{a}\right)=0,\hspace{1.5cm}  \textbf{gh}\left[\bar{\omega}^{a}\right]=\textbf{gh}\left[\bar{\pi}^{a}\right]=0.
\ee
It is necessary to include the antifields associated with these new auxiliary fields with the following properties
\begin{align}
&\epsilon\left(\bar{\xi}^{\star a}\right)=\epsilon\left(\bar{\eta}^{\star a}\right)= 0,  &&\textbf{gh}\left[\bar{\xi}^{\star a}\right]=\textbf{gh}\left[\bar{\eta}^{\star a}\right]=0, \no \\
&\epsilon\left(\bar{\omega}^{\star a}\right)=\epsilon\left(\bar{\pi}^{\star a}\right)=1, &&\textbf{gh}\left[\bar{\omega}^{\star a}\right]=\textbf{gh}\left[\bar{\pi}^{\star a}\right]=-1.
\end{align}
The minimal set together with these new auxiliary fields constitute the so-called \textit{non-minimal} set. The non-minimal extension of BV action reads
\be
S_{BV-nm}=S_{BV}+\bar{\xi}^{\star a} \bar{\pi}^a+ \bar{\eta}^{\star a} \bar{\omega}^a
\ee
By employing the Gaussian-averaging gauge-fixing procedure we have
\begin{equation}
\Psi_{A}^{\star}=\frac{\partial\Theta}{\partial\Psi^{A}}.\label{eq:eliminate_g}
\end{equation}
With this choice we can eliminate the antifields via Eqs. (\ref{eq:gf}) and
(\ref{eq:eliminate_g})
\begin{align}
& A_{\mu}^{\star a} =  -\partial^{\mu}\bar{\xi}^{a}, && \bar{\eta}^{\star a}  =-\frac{\bar{\omega}^a}{2\gamma^\prime} +\partial^{\mu}\phi_{\mu}^{a}, \no \\
&\phi_{\mu}^{\star a} = -\partial^{\mu}\bar{\eta}^{a}, && \xi^{\star a}  = 0, \no \\
&\bar{\xi}^{\star a} = -\frac{\bar{\pi}^a}{2\gamma}+\partial^{\mu}A_{\mu}^{a}, && \eta^{\star a} = 0,\no \\
&\rho^{\star a} = 0.
\end{align}
Finally we obtain the gauge-fixed quantized-ready action for extended JP model
\begin{align} \label{bv-fixed}
S_\Theta  = & S_{0}-\int d^3x \: \Bigg(\partial_{\mu}\bar{\xi}^{a}D^{\mu ab}\xi^{b}-\partial_{\mu}\bar{\eta}^{a}\left(f^{abc}\phi^{\mu b}\xi^{c}+D^{\mu ab}\eta^{b}\right) \no \\
+&  \bar{\pi}^{a}\(-\frac{\bar{\pi}^a}{2\gamma}+\partial^{\mu}A_{\mu}^{a}\) +\bar{\omega}^{a}\(-\frac{\bar{\omega}^a}{2\gamma^\prime} +\partial^{\mu}\phi_{\mu}^{a}\) \Bigg)
\end{align}
The Gaussian integration over auxiliary fields $\bar{\pi}$ and $\bar{\omega}$ can be performed for Eq. (\ref{bv-fixed}) to give
\begin{align}
S_\Theta \longrightarrow & -\frac{1}{4}\int d^{3}x\Big(\frac{1}{2}F^{a\mu\nu}F_{\mu\nu}^a+\frac{1}{2}\left(G^{a\mu\nu}-i\left[F^{\mu\nu},\rho\right]^a\right)\left(G_{\mu\nu}^a-i\left[F_{\mu\nu},\rho\right]^a\right)-m\epsilon^{\mu\nu\rho}F_{\mu\nu}^a\phi_{\rho}^a \no \\
& +\partial_{\mu}\bar{\xi}^{a}D^{\mu ab}\xi^{b} -\partial_{\mu}\bar{\eta}^{a}\left(f^{abc}\phi^{\mu b}\xi^{c}+D^{\mu ab}\eta^{b}\right) 
+  \frac{\gamma}{2}\partial^{\mu}A_{\mu}^a\partial^{\nu}A_{\nu}^a  +\frac{\gamma^\prime}{2}\partial^{\mu}\phi_{\mu}^a\partial^{\nu}\phi_{\nu}^a\Big)
\end{align}
which is very similar to the Yang-Mills action fixed in the $R_\gamma$ gauge. The case $\gamma=\gamma^\prime=1$ is the Feynman gauge. When $\gamma,\gamma^\prime \rightarrow \infty$, the $\bar{\pi}$ and $\bar{\omega}$ dependence in $\Theta$ of Eq. (\ref{eq:gf}) 
disappears and the Landau gauge $\partial^{\mu}\phi_{\mu}^a = \partial^{\mu}A_{\mu}^a=0$ is imposed as a delta-function condition. 

The gauge-fixed BRST transformations are
\begin{align}
&\delta_{B_{\Theta}}A^a_\mu=D_{\mu}^{ab}\xi^b, &&\delta_{B_{\Theta}}\phi^a_\mu=f^{abc}\phi^b_\mu \xi^c+D_\mu^{ab} \eta^b, \no \\
&\delta_{B_{\Theta}}\rho^a=f^{abc}\rho^b\xi^c-\eta^a, &&\delta_{B_{\Theta}}\xi^a=f^{abc}\xi^b\eta^c, \no \\
&\delta_{B_{\Theta}}\eta^a=f^{abc}\xi^b\eta^c, && \delta_{B_{\Theta}}\bar{\xi}^a=\bar{\pi}^a, \no \\
&\delta_{B_{\Theta}}\bar{\eta}^a=\bar{\omega}^a, && \delta_{B_{\Theta}}\bar{\pi}^a=0, \no \\
&\delta_{B_{\Theta}}\bar{\omega}^a=0.
\end{align}
The nilpotency of $\delta_{B_{\Theta}}$ holds off-shell because the original gauge algebra is closed.

The next step would be to discuss the anomalies of this theory and also to calculate its perturbative expansion and anomalies using the above action.  Using this result, we can compare the anomaly given by the NC JP action through the computation made in \cite{Banerjee:2005zq}.  It is an ongoing research.  

As we have mentioned before, for the BV quantization of the non-extended version of the JP model, Eq. (1), we have to consider the value $\rho=0$ for the action and its symmetries.  The field-antifield quantization can be easily obtained for the non-extended model in this way.

%%%%%%%%%%%%%%%%%%%%%%%%%%%%%%%%%%%%%%%%%%%%
%%%%%%%%%%%%%%%%%%%%%%%%%%%%%%%%%%%%%%%%%%%%%%
\section{Noncommutative Jackiw-Pi model}

The NC version of original JP model will be written as
\be \label{nc_jp}
\hat{\mathcal{S}}=Tr\int d^{3}x\left\{ \frac{1}{2}\hat{F}^{\mu\nu}\star\hat{F}_{\mu\nu}+\frac{1}{2}\hat{G}^{\mu\nu}\star\hat{G}_{\mu\nu}-m\epsilon^{\mu\nu\rho}\hat{F}_{\mu\nu}\star\hat{\phi}_{\rho}\right\}.
\ee
In similarity with commutative spacetime, the following definitions in the NC space-time are claimed as
\be
\hat{F}_{\mu\nu}=\partial_{\mu}\hat{A}_{\nu}-\partial_{\nu}\hat{A}_{\mu}-i[\hat{A}_{\mu},\hat{A}_{\nu}]_{\star}
\ee
\be
\hat{G}_{\mu\nu}=\hat{D}_{\mu}\hat{\phi}_{\nu}-\hat{D}_{\nu}\hat{\phi}_{\mu}
\ee
\be
\hat{D}_{\mu}\hat{\phi}_{\nu}=\partial_{\mu}\hat{\phi}_{\nu}-i\hat{[A_{\mu}},\hat{\phi}_{\nu}]_{\star}
\ee

\ni where $[A,B]_{\star}=A\star B-B\star A$ as before. By using the definition of MW star product, up to the first order, we have
\be
[A,B]_{\star}=[A,B]+\frac{i}{2}\theta^{ij}\{\partial_{i}A,\partial_{j}B\}.
\ee
It is worthy to mention again that in a general NC space-time the objects inside the above anticommutator take value in the universal enveloping algebra, $\mathcal{U}\(su(n)\)$.

According to the SW map the gauge transformations are form-invariant, just the fields and operators must be reformulated in NC spacetime. In the other words

\begin{equation}
\left\{ \begin{array}{c}
\delta_{\theta}\hat{A}_{\mu}=\hat{D}_{\mu}\hat{\theta}=\partial_{\mu}\hat{\theta}-i[\hat{A}_{\mu},\hat{\theta}]_{\star}\\
\delta_{\theta}\hat{\phi}_{\mu}=-i[\hat{\phi}_{\mu},\hat{\theta}]_{\star}
\end{array}\right.\label{eq:gg1}
\end{equation}

\ni and

\begin{equation}
\left\{ \begin{array}{c}
\delta_{\chi}\hat{A}_{\mu}=0\\
\delta_{\chi}\hat{\phi}_{\mu}=\hat{D}_{\mu}\hat{\chi}=\partial_{\mu}\hat{\chi}-i[\hat{A}_{\mu},\hat{\chi}]_{\star}
\end{array}\right..\label{eq:gg2}
\end{equation}

The action has three parts that must be mapped to commutative spacetime. The Yang-Mills term, dynamical/interaction term of $\phi_\mu$ and the third one is a Chern-Simons like term. 
As we saw earlier, the SW map gives us a way to express the variables of NC spacetime in terms of commutative ones up to some freedom. Mapping of the Yang-Mills, term up to the first order, is driven by integration of relation (\ref{nc_ff}) and the result is \cite{Jurco:2001rq}

\begin{eqnarray} \label{nc_jp1}
\frac{1}{2}Tr\int\hat{F}^{\mu\nu}\star\hat{F}_{\mu\nu}d^{3}x&=&\frac{1}{2}Tr\int\hat{F}^{\mu\nu}\hat{F}_{\mu\nu}d^{3}x \\
&=&\frac{1}{2}Tr\int d^{3}x \( F^{\mu\nu}F_{\mu\nu}-\frac{1}{2}\theta^{kl}F_{kl}F_{\mu\nu}F^{\mu\nu}+\theta^{kl}F_{\mu k}F_{\nu l}F^{\mu\nu}\). \label{eq:g1-1} \nonumber 
\end{eqnarray}

The vector field $\phi_\mu$ transforms in adjoint representation of the gauge group. So the SW map tells us that, up to the first order, this field can be expressed as
\bn
\hat{\phi}_{\mu} &=&\phi_{\mu} -\frac{1}{4}\theta^{\rho\sigma}\{A_{\rho},\partial_{\sigma}\phi_{\mu}+D_{\sigma}\phi_{\mu}\} \no \\
&\equiv& \phi_\mu+\theta \phi^1_\mu
\en
where $D_{\mu}  \bullet=  \partial_{\mu}\bullet-i[A_{\mu},\bullet]$. 

The second term of action (\ref{nc_jp}) is more complicated and needs more attention. Using the SW map this term can be written as
\bn \label{sw-jp1}
\frac{1}{2}Tr\int d^{3}x \:\hat{G}^{\mu\nu}\star\hat{G}_{\mu\nu}&=&\frac{1}{2}Tr\int d^{3}x\big(\hat{D}^{\mu}\hat{\phi}^{\nu}-\hat{D}^{\nu}\hat{\phi}^{\mu}\big) \star \big(\hat{D}_{\mu}\hat{\phi}_{\nu}-\hat{D}_{\nu}\hat{\phi}_{\mu}\big) \\
&=&\frac{1}{2}Tr\int d^{3}x\big(\hat{D}^{\mu}\phi^{\nu}+\hat{D}^{\mu}\phi^{1\nu}-\hat{D}^{\nu}\phi^{\mu}-\hat{D}^{\nu}\phi^{1\mu}\big) \no \\&& \quad \quad  \qquad \star \big(\hat{D}_{\mu}\phi_{\nu}+\hat{D}_{\mu}\phi_{\nu}^{1}-\hat{D}_{\nu}\phi_{\mu}-\hat{D}_{\nu}\phi_{\mu}^{1}\big). \no
\en
The covariant derivative in the above expression is given by
\be
\hat{D}_{\mu}\phi_{\nu}=D_{\mu}\phi_{\nu}-i\big[A_{\mu}^{1},\phi_{\nu}\big]+\frac{\theta^{\alpha\beta}}{2}\big\{\partial_{\alpha}A_{\mu},\partial_{\beta}\phi_{\nu}\big\}
\ee
where $A^1_\mu$ is the first term of the expansion of NC field $\hat{A}_\mu$ in terms of commutative fields, as we saw in Eq. (\ref{sw-a1}). By plugging in the expanded covariant derivative in Eq. (\ref{sw-jp1}) we obtain
\bn
&&\frac{1}{2}Tr\int d^{3}x(\hat{G}^{\mu\nu})\star(\hat{G}_{\mu\nu}) \\
&=&\frac{1}{2}Tr\int d^{3}x\big(D^{\mu}\phi^{\nu}D_{\mu}\phi_{\nu}-i D^{\mu}\phi^{\nu}\big[A_{\mu}^{1},\phi_{\nu}\big]+\frac{\theta^{\alpha\beta}}{2}D^{\mu}\phi^{\nu}\big\{\partial_{\alpha}A_{\mu},\partial_{\beta}\phi_{\nu}\big\}+ D^{\mu}\phi^{\nu}D_{\mu}\phi_{\nu}^{1} \no \\
&-& D^{\mu}\phi^{\nu}D_{\nu}\phi_{\mu}-i D^{\mu}\phi^{\nu}\big[A_{\nu}^{1},\phi_{\mu}\big]+\frac{\theta^{\alpha\beta}}{2}D^{\mu}\phi^{\nu}\big\{\partial_{\alpha}A_{\nu},\partial_{\beta}\phi_{\mu}\big\}- D^{\mu}\phi^{\nu}D_{\nu}\phi_{\mu}^{1}\no \\
&+& D^{\mu}\phi^{1\nu}D_{\mu}\phi_{\nu}- D^{\mu}\phi^{1\nu}D_{\nu}\phi_{\mu} -i\big[A^{1\mu},\phi^{\nu}\big]D_{\mu}\phi_{\nu} +i\big[A^{1\mu},\phi^{\nu}\big]D_{\nu}\phi_{\mu} \no \\
&+&\frac{\theta^{\alpha\beta}}{2}\big\{\partial_{\alpha}A^{\mu},\partial_{\beta}\phi^{\nu}\big\} D_{\mu}\phi_{\nu}-\frac{\theta^{\alpha\beta}}{2}\big\{\partial_{\alpha}A^{\mu},\partial_{\beta}\phi^{\nu}\big\} D_{\nu}\phi_{\mu} \no \\
&-& D^{\nu}\phi^{\mu}D_{\mu}\phi_{\nu}+i D^{\nu}\phi^{\mu}\big[A_{\mu}^{1},\phi_{\nu}\big]- \frac{\theta^{\alpha\beta}}{2}D^{\nu}\phi^{\mu} \big\{\partial_{\alpha}A_{\mu},\partial_{\beta}\phi_{\nu}\big\}- D^{\nu}\phi^{\mu}D_{\mu}\phi_{\nu}^{1} \no \\
&+& D^{\nu}\phi^{\mu}D_{\nu}\phi_{\mu}+ i D^{\nu}\phi^{\nu} \big[A_{\nu}^{1},\phi_{\mu}\big]- \frac{\theta^{\alpha\beta}}{2}D^{\nu} \phi^{\mu}\big\{\partial_{\alpha}A_{\nu},\partial_{\beta}\phi_{\mu}\big\}+  D^{\nu}\phi^{\mu}D_{\nu}\phi_{\mu}^{1}. \no
\en
After doing some algebra the above expression can be simplified as
\bn 
&&\frac{1}{2}Tr\int d^{3}x(\hat{G}^{\mu\nu})\star(\hat{G}_{\mu\nu})= \\
&&\frac{1}{2}Tr\int d^{3}x\:\Big(G^{\mu\nu}G_{\mu\nu}+3G_{\mu\nu}^{1}D^{\mu}\phi^{\nu}-i G_{\mu\nu}\big[A^{1\mu},\phi^{\nu}\big]+\frac{\theta^{\alpha\beta}}{2}G^{\mu\nu}\big\{\partial_{\alpha}A_{\mu},\partial_{\beta}\phi_{\nu}\big\}\Big) \no
\en
where $G_{\mu\nu}^{1}=D_{\mu}\phi_{\nu}^{1}-D_{\nu}\phi_{\mu}^{1}$. The above expression can be rewritten solely in terms of ordinary fields of commutative theory,
\bn \label{nc_jp2}
&&\frac{1}{2}Tr\int d^{3}x (\hat{G}^{\mu\nu})\star(\hat{G}_{\mu\nu})  \no \\
&=&  \frac{1}{2}Tr\int d^{3}x\:\Big[G^{\mu\nu}G_{\mu\nu}  -3\theta^{\rho\sigma}G^{\mu\nu}\Big(D_{\mu}A_{\rho}\(\partial_{\sigma}+D_\sigma\)\phi_{\nu} -\frac{1}{3}\partial_{\alpha}A_{\mu}\partial_{\beta}\phi_{\nu}\Big) \Big]. 
\en
According to the SW map, the Chern-Simons like term can be transformed as{}
\bn 
mTr\int d^{3}x\:\epsilon^{\mu\nu\rho}\hat{F}_{\mu\nu}\star\hat{\phi}_{\rho}&=& mTr\int d^{3}x\:\epsilon^{\mu\nu\rho}\hat{F}_{\mu\nu}\hat{\phi}_{\rho} \\
&=& mTr\epsilon^{\mu\nu\rho}\int d^{3}x\:\big(F_{\mu\nu}+ F_{\mu\nu}^{1}\big)\big(\phi_{\rho}+\phi_{\rho}^{1}\big) \no \\
&=& mTr\epsilon^{\mu\nu\rho}\int d^{3}x\:\big(F_{\mu\nu}\phi_{\rho}+F_{\mu\nu}^{1}\phi_{\rho}+ F_{\mu\nu}\phi_{\rho}^{1}\big). \no
\en
This expression can also be rewritten just in terms of variables of the original theory
\bn \label{nc_jp3}
&&mTr\int d^{3}x\:\epsilon^{\mu\nu\rho}\hat{F}_{\mu\nu}\star\hat{\phi}_{\rho} \nonumber \\
& = & mTr\epsilon^{\mu\nu\rho}\int d^{3}x\:\Big[F_{\mu\nu}\phi_{\rho}
 +  \theta^{\alpha\beta}\big(F_{\mu\alpha}F_{\nu\beta}\phi_{\rho}+\frac{1}{4}F_{\mu\nu}\big\{\phi_{\rho},(\partial_{\beta}+D_{\beta})A_{\alpha}\big\}\big)\Big]. 
\en

The NC JP theory is given by adding up Eqs.(\ref{nc_jp1}), (\ref{nc_jp2}) and (\ref{nc_jp3})
\bn
\hat{\mathcal{S}}&=& Tr\int d^{3}x\left\{ \frac{1}{2}\hat{F}^{\mu\nu}\star\hat{F}_{\mu\nu}+\frac{1}{2}\hat{G}^{\mu\nu}\star\hat{G}_{\mu\nu}-m\epsilon^{\mu\nu\rho}\hat{F}_{\mu\nu}\star\hat{\phi}_{\rho}\right\} \\
&=&\mathcal{S}+\frac{1}{2}Tr\int d^{3}x \Big( -\frac{1}{2}\theta^{\alpha\beta}F_{\alpha\beta}F_{\mu\nu}F^{\mu\nu}+\theta^{\alpha\beta}F_{\mu \alpha}F_{\nu \beta}F^{\mu\nu} \no \\
&-& 3\theta^{\rho\sigma}G^{\mu\nu}\Big(D_{\mu}A_{\rho}\(\partial_{\sigma}+D_\sigma\)\phi_{\nu} -\frac{1}{3}\partial_{\alpha}A_{\mu}\partial_{\beta}\phi_{\nu}\big) \no \\
& + & \theta^{\alpha\beta}\big(F_{\mu\alpha}F_{\nu\beta}\phi_{\rho}+\frac{1}{4}F_{\mu\nu}\big\{\phi_{\rho},(\partial_{\beta}+D_{\beta})A_{\alpha}\big\}\big)\Big) \no
\en
\ni where ${\cal S}$ is the standard JP model form $\theta=0$, i.e., $\hat{\cal S}(\theta=0)={\cal S}$.

This complete $\mathcal{O}\(\theta^1\)$ noncommutative JP theory contains vertices, with a higher number of gauge bosons, that are absent in the original theory and from the phenomenological point of view these two Lagrangian produce different interactions. We have not included explicitly the structure constants in our analysis so one can not discuss the perturbation expansion of the NC theory. For a future work, we are going to add fermionic matter field in the theory with explicit structure constants and we will analyze its perturbative expansion and the phenomenological aspects of both theories.

\bigskip \bigskip
\ni {\bf \large{The NC version of the extended JP model}}

\bigskip \bigskip

For the extended version of JP action, given by Eq. (\ref{extended}), the introduction of the MW version product is given by

\bn
\label{aa}
\mathcal{L}=tr\int d^{3}x\left\{ \frac{1}{2}\hat{F}^{\mu\nu}\star\hat{F}_{\mu\nu}+\frac{1}{2}\left(\hat{G}^{\mu\nu}-i\left[\hat{F}^{\mu\nu},\hat{\rho}\right]_{\star}\right)\star\left(\hat{G}_{\mu\nu}-i\left[\hat{F}_{\mu\nu},\hat{\rho}\right]_{\star}\right)-m\epsilon^{\mu\nu\rho}\hat{F}_{\mu\nu}\star\hat{\phi}_{\rho}\right\} \no \\ 
\en

\ni where $[A,B]_*\,=\,A*B\,-\,B*A$.  For a non-Abelian theory, the commutator is given by
\bn
[A,B]_* \,=\,[A,B]\,+\,\frac i2 \theta^{ij}\,\{\p_i A , \p_i B \}
\en

and

\be \label{bb}
\hat{F}_{\mu\nu}=\partial_{\mu}\hat{A}_{\nu}-\partial_{\nu}\hat{A}_{\mu}-i[\hat{A}_{\mu},\hat{A}_{\nu}]_{\star}
\ee

\be \label{cc}
\hat{G}_{\mu\nu}=\hat{D}_{\mu}\hat{\phi}_{\nu}-\hat{D}_{\nu}\hat{\phi}_{\mu}
\ee

\ni where
\be \label{cc1}
\hat{D}_{\mu}\hat{\phi}_{\nu}=\partial_{\mu}\hat{\phi}_{\nu}-i\hat{[A_{\mu}},\hat{\phi}_{\nu}]_{\star}
\ee

According to Seiberg-Witten map the gauge transformations form does
not change, just the fields and operators must be transformed into
NC space. In other word 

\begin{equation} 
\left\{ \begin{array}{c}
\delta_{\theta}\hat{A}_{\mu}=\hat{D}_{\mu}\hat{\theta}=\partial_{\mu}\hat{\theta}-i[\hat{A}_{\mu},\hat{\theta}]_{\star}\\
\delta_{\theta}\hat{\phi}_{\mu}=-i[\hat{\phi}_{\mu},\hat{\theta}]_{\star}\\
\delta_{\theta}\hat{\rho}=-i[\hat{\rho},\hat{\theta}]_{\star}
\end{array}\right.\label{eq:gg1}
\end{equation}

\noindent and

\begin{equation} 
\left\{ \begin{array}{c}
\delta_{\chi}\hat{A}_{\mu}=0\\
\delta_{\chi}\hat{\phi}_{\mu}=\hat{D}_{\mu}\hat{\chi}=\partial_{\mu}\hat{\chi}-i[\hat{A}_{\mu},\hat{\chi}]_{\star}\\
\delta_{\chi}\hat{\rho}=-\hat{\chi}
\end{array}\right..\label{eq:gg2}
\end{equation}

\ni where (\ref{eq:gg1}) is the first set of gauge transformation, which is more interesting than the second one since $A_\mu$ transform.   
Then, we believe that, since the first set of gauge symmetries presents a non-trivial gauge transformation for $A_\mu$, the NC version of (\ref{extended}) will bring interesting extra terms connected to this field.
In this way we will provide the NC version for this first set of gauge transformation, for the second set, the calculation is analogous and easier than the one we present here.

After a huge algebraic work, using the cyclic properties of the trace, the definitions above, the NC version of the JP model considering the first set of gauge transformations, Eq. (\ref{eq:gg1}), is
\begin{align}
&\mathcal{L}_{NC}=\mathcal{L}+tr\int d^{3}x\bigglb\{-\frac{h}{8}\theta^{kl}F_{kl}F_{\mu\nu}F^{\mu\nu}+\frac{h}{2}\theta^{kl}F_{\mu k}F_{\nu l}F^{\mu\nu} \no \\
&+\frac{h}{2}\theta^{ij}\partial^{\nu}\left(A^{i}\partial^{j}\phi^{\mu}+A^{i}D^{j}\phi^{\mu}\right)(\partial_{\mu}\phi_{\nu}-\partial_{\nu}\phi_{\mu})-\frac{ih}{2}\theta^{ij}\partial_{i}F_{\mu\nu}\partial_{j}\rho(\partial^{\mu}\phi^{\nu}-\partial^{\nu}\phi^{\mu}) \nonumber \\
&-\frac{ih}{2}\theta^{ij}\partial_{i}F^{\mu\nu}\partial_{j}\rho(\partial_{\mu}\phi_{\nu}-\partial_{\nu}\phi_{\mu})+\frac{h}{2}\theta^{ij}\partial_{\nu}\left(A_{i}\partial_{j}\phi_{\mu}+A_{i}D_{j}\phi_{\mu}\right)(\partial^{\mu}\phi^{\nu}-\partial^{\nu}\phi^{\mu}) \nonumber \\
&+mh\epsilon^{\mu\nu\rho}\Big(\theta^{ij}F_{\mu\nu}A_{i}\left(\partial_{j}+D_{j}\right)\phi_{\rho}+\theta^{ij}F_{\mu i}F_{\nu j}\phi_{\rho}+\frac{i}{2}\theta^{ij}\partial_{i}F_{\mu\nu}\partial_{j}\phi_{\rho}\Big)\biggrb\}
\end{align}%

\ni which recovers Eq. (\ref{extended}) when $\theta=0$ and we can see that the extra terms present higher derivatives a mass terms.

%%%%%%%%%%%%%%%%%%%%%%%%%%%%%%%%%%%%%%%%%%%%%%%%%%
%%%%%%%%%%%%%%%%%%%%%%%%%%%%%%%%%%%%%%%%%%%%%%%%%%%%%%
\section{Conclusions and perspectives}

In this work we have discussed the behavior of the JP model under the introduction of Planck scale elements through two different formalism, the NC one and the BV quantization method.  The BV action, the gauge fixed action and the BRST transformations were computed.  The Nakanishi-Lautrup field was introduced.

Concerning NCy, the non-Abelian JP model shows that the MW product and the SW map can be introduced and the final NC version was obtained.  As a further step, we can compute the anomaly of both NC and commutative actions and compare the results with the mapping for NC anomalies developed in \cite{Banerjee:2005zq}.  This is an ongoing research and will be published elsewhere.

%%%%%%%%%%%%%%%%%%%%%%%%%%%%%%%%%%%%%%%%%%%%%%%%%%%%%%%%%%%%%%%%%
%%%%%%%%%%%%%%%%%%%%%%%%%%%%%%%%%%%%%%%%%%%%%%%%%%%%%%%%%%%%%%%%%%%
\appendix

\section{Noncommutative gauge theory}\label{section6}

Gauge theories are crucially important when they can build a realistic physical model and are the main ingredients of standard model of particle physics. So, in order to obtain any real results out of the NC field theory, the notion of gauge symmetry had to be generalized to the NC setting. Since gauge symmetries are essentially local, generalizing them to the nonlocal NC spacetime is highly nontrivial.\\

There are two methods to construct gauge field theories in NC spacetime. The first one uses the SW map, obtained from string theory \cite{sw}, which maps a NC gauge theory into a commutative gauge theory. The second one uses a NC generalization of a gauge group and the $\star$-product to construct a gauge theory in the framework of NC field theory. Both methods have been further developed and they offer some flexibility in their approaches. In this chapter we shall study just the SW method briefly in the case of the constant $\theta$ and then we will construct a NC version of a non-Yang-Mills gauge theory with $SU(N)$ gauge group. The reader with an interest in field theoretical approach can refer to \cite{Chaichian:2001py, Chaichian:2009uw,Chaichian:2007tk,Chaichian:2008ge,Chaichian:2004za}.\\

Until now we have discussed Lorentz-invariant NC spacetime where the NC parameter is an operator valued object, but now we will take a look at the cases where the NC space-time is considered to be the canonical one, i.e. the NC parameter is a real valued \textit{constant} matrix. In this type of noncommutativity (NCy) the Lorentz invariance is violated.\\

For the future use the Moyal-Weyl (MW) $\star$-product and the Moyal bracket (For a review of MW product, see \cite{reviews1,reviews2}) are naturally generalized for the algebra of matrix-valued functions $M_{n \times n} \otimes A_\theta$, i. e., for two arbitrary functions $f(x)$ and $g(y)$ we have
\be
\left( f(x) \star g(y)\right)_{ij} = f(x)_{ik} \star g(y)_{kj}.
\ee
The Hermitian conjugation for the algebra $M_{n \times n} \otimes A_\theta$ can be defined by the usual Hermitian conjugation of matrices 
$\left( f(x)^\dagger \right)_{ij} = \left( f(x)_{ji}^\star \right)$ and by the definition that the $\star$-product behaves under the operation
\be
\left( f(x) \star g(x)  \right)^\dagger = g(x)^\dagger \star f(x)^\dagger.
\ee

\ni in the next section we will talk more about the MW product and it we will show it explicitly.

\subsection{The Seiberg-Witten map and universal enveloping algebra}

After a quantization process, the open string theory in a constant antisymmetric background field,
with string end points constrained on D-branes, by using the Pauli-Villars and the point-splitting regularization, one obtains a commutative or NC gauge theory, respectively. The SW map provides a correspondence between these two gauge theories, which should be equivalent, since a well-defined quantum theory does not depend on the regularization technique.

The SW map, as originally proposed, is a map between the NC $U_\star(N)$ gauge theory, described by $\hat{A}$ and $\Lambda$ as gauge field and the gauge transformations, respectively and the corresponding ordinary commutative $u(N)$-matrix valued functions $A$ and $\Lambda$. In this approach it is argued that, because most of the gauge theories on NC spaces cannot be constructed with Lie algebra valued infinitesimal gauge transformations, the infinitesimal gauge transformations should instead, be taken to be enveloping algebra valued. The idea is to bypass the difficulties in constructing NC gauge groups by letting the generators of the gauge transformations and the gauge fields to take values in the universal enveloping of the corresponding gauge algebra. The main problem with this approach is that enveloping algebras are infinite dimensional, which means that simply the numbers of both gauge transformation parameters and the gauge fields are infinite.

The gauge transformation parameters and the gauge fields can, however, be defined to be functions of the corresponding Lie algebra valued objects, i.e., the functions being obtained through the SW maps, so that their numbers are the same as in the corresponding commutative gauge theories.

Let us consider the NC version of a gauge theory of a generic non-Abelian gauge algebra, say the algebra $su(n)$, with the matter fields $\hat{\psi}$ and the gauge fields $\hat{A}_\mu$. The infinitesimal local gauge transformations are
\bn
\hat{\delta}_{\hat{\Lambda}} \hat{\psi} &=& i \rho_\psi(\hat{\Lambda} (x)) \star \hat{\psi} \label{nc-matter}\\
\hat{\delta}_{\hat{\Lambda}} \hat{A}_\mu &=& \hat{\p}_\mu \hat{\Lambda}({x}) + i \left[\hat{\Lambda}({x}) , \hat{A}_\mu\right]_\star
\en
where the NC infinitesimal gauge transformation parameter $\hat{\Lambda}$ is valued in a universal enveloping of the gauge algebra $\mathcal{U}(su(n))$ and $\rho_\psi$ is the matter representation of $\mathcal{U}(su(n))$. It should be noted that there is no gauge symmetry group, since this gauge symmetry is only defined for infinitesimal gauge transformations\footnote{For a $\mathcal{U}(L)$ there is nothing like the exponential map that maps a Lie algebra $L$ to a Lie group.}. Generally speaking, the gauge transformation parameter $\hat{\Lambda}$ cannot be Lie algebra valued, because the commutator of two Lie algebra valued parameters $\hat{\Lambda}=\hat{\Lambda}_i T_i$ and $\hat{\Sigma} = \hat{\Sigma}_i T_i$ does not close in the Lie algebra with the gauge transformations
\be
\left[ \hat{\Lambda} \: , \: \hat{\Sigma}\right]_\star= \frac{1}{2} \{\hat{\Lambda}_i \: , \: \hat{\Sigma}_j\}_\star \underbrace{\[T_i \: , \: T_j\]}_{if_{ijk}T_k}  + \frac{1}{2}\underbrace{\[ \hat{\Lambda}_i \: , \: \hat{\Sigma}_j\]_\star}_{\neq 0} \{T_i \: , \: T_j \}.
\ee

Therefore, we have to use the fields and gauge transformations that are $\mathcal{U}(su(n))$-valued.
The gauge fields $\hat{A}_\mu$ have to be in the adjoint representation of $\mathcal{U}(su(n))$. The gauge covariant derivative and the field strength are given by

\bn
\hat{D}_\mu \hat{\psi} &=& \p_\mu \hat{\psi} - i \rho_\psi(\hat{A}_\mu) \star \hat{\psi} \\
\hat{F}_{\mu\nu} &=& \p_{[\mu} \hat{A}_{\nu ]} - i \[\hat{A}_\mu \: , \: \hat{A}_\nu\]_\star
\en

\noindent with the gauge transformations

\bn
\hat{\delta}_{\hat{\Lambda}} \hat{D}_\mu \hat{\psi} &=& i\hat{\Lambda}(x) \star \hat{D}_\mu  \hat{\psi} \\
\hat{\delta}_{\hat{\Lambda}} \hat{F}_{\mu\nu} &=& i \[\hat{\Lambda}(x) , \hat{F}_{\mu\nu}\]_\star.
\en

The gauge invariant action for the gauge sector is defined by
\be
S\left[\hat{A}, \p\hat{A}\right] = -\frac{1}{4}\int d^Dx \: \text{Tr} \left(\hat{F}_{\mu\nu} \hat{F}^{\mu\nu}\right)
\ee
and the action for the matter/interaction sector is constructed by using the covariant derivative.
For example, the action of a NC fermion is written as
\be
S\left[ \hat{\psi}, \p\hat{\psi}, \hat{A}\right] = \int d^dx \: \bar{\hat{\psi}} \star (\gamma^\mu \hat{D}_\mu - m)\hat{\psi}.
\ee
These definitions are similar to the corresponding commutative $su(n)$ gauge theory, the differences being the ordinary point-wise product and the Lie algebra valued fields and the gauge transformations parameters. Here we denote the commutative concepts without the hats: $\psi, A_\mu, \Lambda$ etc. In order to fix the notation we mention that in the commutative space, the fields transform under gauge transformations with Lie algebra-valued infinitesimal parameters
\be
\de_{\Lambda}\psi(x) = i \Lambda(x)\psi(x) \quad ; \quad \Lambda(x) = \Lambda_a T^a.
\ee
The commutator of two gauge transformations gives us
\be \label{cm-1}
\(\de_{\Lambda} \de_{\Sigma} - \de_{\Sigma} \de_{\Lambda} \)\psi(x) = i \Lambda_a(x) \Sigma_b(x)f_{abc} T^c \psi(x) = \de_{\Lambda \times \Sigma} \psi(x),
\ee
where

\be
\Lambda \times \Sigma \equiv \Lambda_a \Sigma_b f_{abc}T_c = -i \[\Lambda \: , \: \Sigma\].
\ee

\noindent For the Lie algebra-valued gauge potential $A_{a\mu}(x)$ we define the following transformation
\be
\de_{\Lambda} A_{a\mu} = \p_\mu \Lambda_a - f_{abc} \Lambda_b(x)A_{c\mu}(x) \quad ; \quad A_\mu = A_{a\mu}(x)T_a.
\ee

\noindent Since the gauge invariance of the commutative gauge theory should be maintained in the NC space, the gauge transformations in the latter theory are induced by the transformations of the former theory
\bn
\hat{A}_\mu [A] + \hat{\delta}_{\hat{\Lambda}[\Lambda, A]} \hat{A}_\mu [A] &=& \hat{A}_\mu [A+\delta_{\Lambda} A],\\
\hat{\psi}[\psi,A] + \hat{\delta}_{\hat{\Lambda}[\Lambda, A]} \hat{\psi}[\psi, A] &=& \hat{\psi}[\psi+\delta_{\Lambda} \psi \: , \: A+\delta_{\Lambda} A].
\en
These relations are called SW map. They say that, if the commutative fields $A_\mu$ and $\psi$ are related to the fields $A^{U}_\mu$ and $\psi^{U}$ through the gauge transformation $U=\text{exp}(i\Lambda)$ generated by $\Lambda$, then the NC fields $\hat{A}_\mu[A]$ and $\hat{\psi}[\psi, A]$ are related to the fields $\hat{A}_\mu[A^{U}]$ and $\hat{\psi}[\psi^{U}, A^{U}]$ through the gauge transformation $\hat{U}=\text{exp}(i\hat{\Lambda}[\Lambda, A])$, generated by $\hat{\Lambda}[\Lambda, A]$. These gauge equivalence relations can be solved pertubatively in $\theta$ in order to obtain the SW maps explicitly. For the gauge theories with $U(N)$ as the gauge group, the SW map for the leading order in $\theta$ can be written as

\bn
\hat{A}_\mu \[A\] &=& A_\mu + \frac{1}{4}\theta^{\nu\rho} \{A_\rho \: , \: \p_\nu A_\mu + F_{\mu\nu} \}+ \mathcal{O}\(\theta^2\) \\
\hat{\psi}\[\psi,A\] &=& \psi + \frac{1}{2}\theta^{\mu\nu}\rho_\psi(A_\nu) \p_\mu \psi + \frac{i}{8}\theta^{\mu\nu}\[\rho_\psi(A_\mu) \: , \: \rho_\psi(A_\nu)\]\psi + \mathcal{O}\(\theta^2\)\\
\hat{\Lambda}\[\Lambda , A\] &=& \Lambda + \frac{1}{4} \theta^{\mu\nu}\{A_\nu \: , \: \p_\mu\Lambda\} + \mathcal{O}\(\theta^2\).
\en
As we have mentioned above, the gauge parameters of a general gauge theory, for example, with $SU(N)$ as the gauge group, in the NC space can not be Lie algebra-valued, because the commutation relation is not always closed, they have to take values in enveloping algebra\footnote{As mentioned above just as in the case of $U(N)$ gauge group, one can find that the commutation is closed and the parameters are Lie algebra-valued.}.

\begin{eqnarray*}
	\hat{\Lambda}(x) & = & \hat{\Lambda}_{a}(x)T^{a}+\hat{\Lambda}_{ab}^{1}(x):T^{a}T^{a}:+\ldots\\
	& + & \hat{\Lambda}_{a_{1}a_{2}...a_{n}}^{n-1}(x):T^{a_{1}}\cdots T^{a_{n}}:+\ldots
\end{eqnarray*}
The dots mean that we must sum over a basis of the vector space spanned
by homogeneous polynomials of generators of the Lie algebra. Completely
symmetrized products form such the following basis

\begin{eqnarray*}
	:T^{a}: & = & T^{a}\\
	:T^{a}T^{b}: & = & \frac{1}{2}\left\{ T^{a},T^{b}\right\} =\frac{1}{2}\left(T^{a}T^{b}+T^{b}T^{a}\right)\\
	:T^{a_{1}}\ldots T^{a_{n}}: & = & \frac{1}{n!}\sum_{\pi\epsilon S_{n}}T^{a_{\pi(1)}}\cdots T^{a_{\pi(n)}}.
\end{eqnarray*}
The $\star$-commutator of two enveloping algebra-valued transformations always will remain enveloping algebra-valued. The bad point is that we will deal with a series of infinite parameters, however it is possible to define a gauge transformation where all these infinitely parameters depend on the usual gauge parameter $\Lambda(x)$, the gauge potential $A_{\mu}(x)$ and their derivatives \cite{Jurco:2001rq}. Transformations of this type will be denoted as $\hat{\Lambda}\[A\]$ and their $x$-dependence is purely via this finite set of parameters and gauge potentials $\Lambda\[A\] \equiv \hat{\Lambda}\[A(x)\]$ (for constant $\theta$).

Now the gauge transformation (\ref{nc-matter}) will take the following form
\be
\delta_{\hat{\Lambda}}\hat{\psi}(x)=i\hat{\Lambda}\left[A\right] \star \hat{\psi}(x).
\ee

Each finite set of parameters $\Lambda^0_a(x)$ defines a tower $\Lambda_{\Lambda^0}\left[A^0\right]$ in the enveloping algebra that is completely determined by the Lie algebra-valued part. To define and construct this tower we demand a similarity with Lie algebra \cite{Jurco:2000ja}
\be
\(\de_{\hat{\Lambda}}\de_{\hat{\Sigma}} - \de_{\hat{\Sigma}} \de_{\hat{\Lambda}} \) \hat{\psi}(x) = \de_{\hat{\Lambda} \times \hat{\Sigma}} \hat{\psi}(x).
\ee 
More explicitly we have
\be \label{nc-demand1}
i \de_{\hat{\Lambda}} \hat{\Sigma}\[A\] - i\de_{\hat{\Sigma}} \hat{\Lambda}\[A\] + \hat{\Lambda}\[A\] \star \hat{\Sigma}\[A\] - \hat{\Sigma}\[A\] \star \hat{\Lambda}\[A\] = i \hat{\Omega}_{\hat{\Lambda} \times \hat{\Sigma}} \[A\].
\ee
Now we can use the expansion of the $\star$-product (the MW product, as we have mentioned before) to solve Eq. (\ref{nc-demand1}) in its NC part. 
\begin{eqnarray*}
	\left(f\star g\right)(x) & = & \text{exp}\left(\frac{i}{2}\frac{\partial}{\partial x^{i}}\theta^{ij}\frac{\partial}{\partial y^{j}}\right)f(x)g(y)|_{y\rightarrow x}\\
	& = & f(x)g(x)+\frac{i}{2}\theta^{ij}\partial_{i}f\partial_{j}g+\cdots.
\end{eqnarray*}

We will assume that always the following expansion is possible

\be
\hat{\Lambda}\left[A\right]=\Lambda+\Lambda^{1}\left[A\right]+\Lambda^{2}\left[A\right]+\cdots.
\ee
This expansion is the main ingredient for the construction of
non-Abelian NC gauge theories. If we substitute the above relation
in (\ref{nc-demand1}) to zeroth order, we yield the Eq. (\ref{cm-1}) which is the commutator of two Lie algebra-valued objects. Concerning the first order, by means of an ansatz, we have that

\be \label{nc-envelop-1}
\Lambda^{1}\[A\]=\frac{1}{4}\theta^{\mu\nu}\left\{ \partial_{\mu}\Lambda,A_{\nu}\right\} = \frac{1}{2} \theta^{\mu\nu} \p_\mu \Lambda_a A_{b\nu} :T^a T^b:.
\ee
Also we can expand the fields and gauge potential in NC space in terms
of the original space ones as follows
\be
\hat{\psi}=\psi^{0}+\psi^{1}+...
\ee
and
\be
\hat{A_{\mu}}=A_{\mu}+A_{\mu}^{1}+....
\ee
By the same treatment as the gauge parameter for the gauge potential and field strength at the first order terms one finds \cite{Jurco:2001rq} that
\be \label{sw-a1}
A_{k}^{1}=-\frac{1}{4}\theta^{ij}\left\{ A_{i},\partial_{j}A_{k}+F_{jk}\right\} \,\,,
\ee
\be
F_{ij}^{1}=\frac{1}{2}\theta^{kl}\left\{ F_{ik},F_{jl}\right\} -\frac{1}{4}\theta^{kl}\left\{ A_{k},\left(\partial_{l}+D_{l}\right)F_{ij}\right\} \,\,.
\ee
Hence the ordinary Yang-Mills term $F_{ij} F^{ij}$ in the NC spacetime takes the following form
\bn \label{nc_ff}
\hat{F}_{ij} \star \hat{F}^{ij} &=& F_{ij} F^{ij} +\frac{i}{2} \theta^{kl} D_k F_{ij} D_l F^{ij} +\frac{1}{2} \theta^{kl} \{\{F_{ik},F_{jl}\},F^{ij}\} \no \\ 
&-& \frac{1}{4} \theta^{kl}\{F_{kl}, F_{ij}F^{ij}\} - \frac{i}{4} \theta^{kl} \left[A_k, \{A_l , F_{ij}F^{ij}\}\right].
\en
For the matter field in the fundamental representation we have that

\be
\psi^{1}=-\frac{1}{4}\theta^{ij}A_{i}\left(\partial_{j}+D_{j}\right)\psi  \quad \text{where} \quad D_i \psi = \p_i \psi - iA_i \psi
\ee

\ni and in the adjoint representation \cite{Ulker:2007fm} 

\be
\psi^{1}=-\frac{1}{4}\theta^{ij}\left\{ A_{i},\left(\partial_{j}+D_{j}\right)\psi\right\}  \quad \text{where} \quad D_i \psi = \p_i \psi -i \[A_i , \psi \]
\ee
We must take care that these variables do not take value in a Lie algebra
but in an enveloping algebra. So $\{\bullet,\bullet\}$ is not the anticommutator of a Lie algebra-valued matrices and the result is more complicated such as the one 
in Eq. (\ref{nc-envelop-1}).

The higher order of expansions are obtained analogously. In \cite{Jurco:2001rq} the action of a NC gauge theory with fermionic matter has been constructed to the second order of NCy parameter $\theta$.  The result can be written solely in terms of the usual gauge covariant derivatives and field strengths, which exhibits beautifully the usual gauge invariance of the expansion.

\subsection{The no-go theorem} 

In a realistic physical model we need to consider gauge groups with
several simple factors. Let $G_1$ and $G_2$ be two local gauge groups. The gauge group
$G=G_1 \times G_2$ is defined by
\bn \label{no-go}
g=g_1 \times g_2 \quad ; \quad h=h_1 \times h_2 \quad ; \quad g,h\in G \quad ; \quad g_i,h_i \in G_i \no \\
g.h= \(g_1 \times g_2\) . \(h_1 \times h_2\) \equiv \(g_1.h_1\) \times \(g_2 . h_2\).
\en

\ni where ``·" $\:$ is the corresponding group multiplication for each group. If we now take the
groups to be the NC ones, $G_1= U_\star\(n\)$ and $G_2= U_\star\(m\)$, we can see that, 
because of the $\star$-product we cannot re-arrange the elements of the subgroups as
in (\ref{no-go}). Therefore the matter fields cannot be in the fundamental representation
of both $U_\star\(n\)$ and $U_\star\(m\)$. However, there is one possibility left. The matter field
$\Psi$ can be in the fundamental representation of one group, say $U_\star\(n\)$, and in the
anti-fundamental representation of the other group
\be
\Psi \longrightarrow \Psi' = U \star \Psi \star V^{-1} \quad ; \quad U\in U_\star\(n\), V\in U_\star\(m\).
\ee
In the general case the gauge group consists of N factors $G=\prod^N_{i=1} U_\star\(n_i\)$. The
matter fields can at most be charged under two of the $U_\star\(n_i\)$ factors and they have
to be singlets under the rest of them. This is a strong constraint on the possible models specially the extension of the standard model of particle physics on NC spacetimes.

%%%%%%%%%%%%%%%%%%%%%%%%%%%%%%%%%%%%%%%%%%%%%%
%%%%%%%%%%%%%%%%%%%%%%%%%%%%%%%%%%%%%%%%%%%%%%%%%%%%%%

\section{A Fast Review of Field-Antifield (or Batalin-Vilkovisky) Formalism} \label{section7}

The basic idea of the so-called Field-Antifield formalism is to generalize the BRST invariance to theories with arbitrary gauge structure. The ingredients are the ordinary fields $\Phi^A$, the ghosts, the auxiliary fields and their canonically conjugated antifields $\Phi^\star_A$. With all these elements we can construct the 
well-known field-antifield or Batalin-Vilkovisky (BV) action. At the classical level, the BV action becomes the ordinary classical action when all the antifields are zeroed. A gauge-fixed action can be obtained by a canonical transformation. At this time we can say that the action is in a gauge-fixed basis. The other way to fix the gauge is through the choice of a gauge fermion and to make the antifields to be equal to the functional derivative of this fermionic function.

This method can be applied to gauge theories which have an open algebra (the
algebra of gauge transformations closes only on shell), to closed algebras, to gauge
theories that have structure functions rather than constants (soft algebras), and to
the case where the gauge transformations may or may not be independent, reducible
or irreducible algebras respectively. Zinn-Justin introduced the concept of sources of
BRST-transformations \cite{rollnik1975trends}. These sources are the antifields in the BV formalism. It was shown also that the geometry of the antifields have a natural origin \cite{Witten:1990wb}.

At the quantum level, the FA formalism also works at one-loop anomalies
\cite{Troost:1989cu,DeJonghe:1993zc}. Here, with the addition of extra degrees of freedom, which leads to an extension of the original configuration space, we have a solution for the regularized quantum master equation (QME) at one-loop that has been obtained as an independent part
of the antifields inside the anomaly.

\subsection{Gauge structure}

In a gauge theory the action is invariant under a set of gauge transformations
with infinitesimal form
\be \label{bv-stru}
\delta\Psi^{i}(x)=(R_{\alpha}^{i}\varepsilon^{\alpha})(x) \,\,,
\ee
where $i=1,2,\cdots n$ is the number of fields, $\alpha=1,2,\cdots m<n$
is the number of sets of gauge transformations and $R^i_\alpha$ are the generators of the gauge transformations. The $\varepsilon^{\alpha}$
are the infinitesimal gauge parameters and $R_{\alpha}^{i}$ the generators
of the gauge transformations. When $\epsilon_{\alpha}=\epsilon\left(\varepsilon^{\alpha}\right)=0$
we have an ordinary symmetry, when $\epsilon_{\alpha}=1$ the equation
is characteristic of a supersymmetry. The Grassmann parity of generators
of the gauge transformations is defined as $\epsilon\left(R_{\alpha}^{i}\right)=\epsilon_{\alpha}+\epsilon_{i}$.
Also we have $\epsilon_{i}=\epsilon\left(\phi^{i}\right)$ that defines the 
Grassmann parity of the fields. Fields with $\epsilon_{i}=0$ are
called bosonic and with $\epsilon_{i}=1$ are fermionic. The relation (\ref{bv-stru}) is written in the DeWitt compact notation and its original form is
\be
\delta\Psi^{i}(x)= \sum_{\alpha}\int dy \: R^i_\alpha(x,y)\epsilon^\alpha(y)
\ee
The graded commutation rule is defined as
\be
\phi^{i}\left(x\right)\phi^{j}\left(y\right)=\left(-1\right)^{\epsilon_{i}\epsilon_{j}}\phi^{j}\left(y\right)\phi^{i}\left(x\right).
\ee
Let $S_{0,i}(\phi,x)$ denote the variation of the action with respect to $\phi^i(x)$:
\be
S_{0,i}\equiv \frac{\p_rS_0[\phi]}{\p\phi^i(x)}
\ee
where the subscript $i$ denotes the right derivative with
respect to the corresponding field, that is, the field is to be commutated
to the far right and then dropped. When using right derivatives, the variation $\delta S_0$, or of any other object, is given by $\delta S_0=S_{0,i}\delta\phi^i$. If one were to use left derivatives, the variation of $S_0$ would be read $\delta S_0=\delta\phi^i\frac{\p_lS_0}{\p \phi^i}$.
The commutation rule for the gauge transformations in the most general form obeys
the following relationship
\begin{equation}
\left[\delta_{1},\delta_{2}\right]\phi^{i}=\left(R_{\gamma}^{i}T_{\alpha\beta}^{\gamma}-S_{0,j}E_{\alpha\beta}^{ij}\right)\varepsilon_{1}^{\beta}\varepsilon_{2}^{\alpha}\label{eq:commug}
\end{equation}
where the tensors $T_{\alpha\beta}^{\gamma}$ are called the structure constants of the gauge algebra, although they depend, in general, on the fields of the theory. When $E_{\alpha\beta}^{ij}=0$,
the gauge algebra is said to be closed, otherwise it is open.
Equation (\ref{eq:commug}) defines a Lie algebra if the algebra is closed
and the $T_{\alpha\beta}^{\gamma}$ are independent of the fields.
We will see that the Jackiw-Pi model has a closed and Lie algebraic gauge
structure. 

When we say that the action is invariant under the gauge transformation in
Eq.(\ref{bv-stru}) means that the Noether identities
\be
\int \: dx \: \sum_{i=1}^{n}S_{0,i}(x)R^i_\alpha(x,y)=0 
\ee
hold, or equivalently, in compact notation
\be
S_{0,i}R^i_\alpha=0.
\ee
Hence the field equations may be written as
\be
S_{0,i}=0.
\ee
As in the familiar Faddeev-Popov procedure, it is useful
to introduce ghost fields $C$ with opposite Grassmann parities to
the gauge parameters $\varepsilon^{\alpha}$
\be
\epsilon\left(C^{\alpha}\right)=\epsilon_{\alpha}+1\left(mod\hspace{1em}2\right)
\ee
and to replace the gauge parameters by ghost fields. 

\subsection{Irreducible and reducible gauge theories}

It is important to know any dependences among the gauge generators. After analyzing these relations it is possible to determine the independent degrees of freedom. The simplest gauge theories, for which all gauge transformations are independent, are called irreducible. When dependences exist, the theory is reducible. In reducible gauge theories, there is a ``kind of gauge invariance for gauge transformations'' or what one might call ``level-one'' gauge invariances. If the level-one gauge transformations are independent, then the theory is called first-stage reducible. This may not happen.
Then, there are ``level-two'' gauge invariances, i.e., gauge invariances for the level-one gauge invariances and so on. This leads to the concept of an L-th stage reducible theory. In what follows we let ``m"  denote the number of gauge generators at the s-th stage regardless of whether they are independent.

In this brief review we will consider only theories with irreducible gauge structure. For more detailed discussion of the full formalism the interested reader is encouraged to see \cite{Gomis:1994he,mhct}. 

\subsection{Introducing the antifields }

We incorporate the ghost fields into the field set $\Psi^{A}=\left\{ \phi^{i},C^{\alpha}\right\} $
, where $i=1,...,n$ and $\alpha=1,...,m$. We call it a $minimal$
set. Clearly $A=1,...,N$, where $N=n+m$. One then further increases
the set by introducing an antifield $\Psi_{A}^{\star}$ for each field
$\Psi^{A}$. The Grassmann parity of the antifields is $\epsilon\left(\Psi_{A}^{\star}\right)=\epsilon\left(\Psi^{a}\right)+1\left(mod\hspace{1em}2\right)$.

We assign a new number to each field, the ghost number \textbf{gh}, which is defined as follow
\begin{eqnarray*}
	\textbf{gh}\left[\phi^{i}\right] & = & 0\\
	\textbf{gh}\left[C^{\alpha}\right] & = & 1\\
	\textbf{gh}\left[\Psi_{A}^{\star}\right] & = & -gh\left[\Psi_{A}\right]-1.
\end{eqnarray*}
In this generalized space, the antibracket is defined by
\be
\left(X,Y\right)=\frac{\partial_{r}X}{\p \Psi^{A}}\frac{\partial_{l}Y}{\p \Psi_{A}^{\star}} -\frac{\partial_{r}X}{\p \Psi_{A}^{\star}}\frac{\partial_{l}Y}{\p \Psi^{A}}
\ee
where $\partial_{r}$ denotes the right derivative and $\partial_{l}$ the left
derivative. The antibracket is graded antisymmetric
\be
\left(X,Y\right)=-\left(-1\right)^{\left(\epsilon_{X}+1\right)\left(\epsilon_{Y}+1\right)}\left(Y,X\right).
\ee
If one groups the fields and the antifields together into the set
\be
z^{a}=\left\{ \Psi_{A}^{\star},\Psi^{A}\right\} \hspace{1em}a=1,2,..,2N
\ee
then the antibracket is seen to define a symplectic structure on the
space of fields and antifields
\be
\left(X,Y\right)=\frac{\partial_{r}X}{\partial z^{a}}\omega^{ab}\frac{\partial_{l}Y}{\partial z^{b}}
\ee
with
\be
\omega^{ab}=\left(\begin{array}{cc}
	0 & \delta_{B}^{A}\\
	-\delta_{B}^{A} & 0
\end{array}\right).
\ee
The antifield can be thought of as a kind of conjugate variable to the field,
since
\be
\left(\Psi^{A},\Psi_{B}^{\star}\right)=\delta_{B}^{A}.
\ee
As it can be seen the antibracket is, in some sense, very similar to the Poisson bracket in the phase-space. In fact, by introducing the antifields and defining the antibracket we have an odd(even) symplectic structure inside the Lagrangian formalism. In this way, we can enjoy the clarity and power of Hamiltonian formalism right inside the extended configuration space. 

The antibracket of two fermionic fields is
\be
\(F,F\)=0,
\ee 
for two bosonic fields is
\be
\(B,B\)= 2\frac{\p B}{\p \Psi^A}\frac{\p B}{\p \Psi^\star_B}
\ee
and for any field $X$, the triple commutation gives
\be
\(X,\(X,X\)\)=0.
\ee

\subsection{The classical master equation}

Let $S\left[\Psi^{A},\Psi_{B}^{\star}\right]$ be a functional
of the fields and antifields with the dimension of an action, vanishing
ghost number and even Grassmann parity. The equation
\begin{equation} 
\left(S,S\right)=2\frac{\partial S}{\partial\Psi^{A}}\frac{\partial S}{\partial\Psi_{A}^{\star}}=0\label{eq:master}
\end{equation}
is the classical master equation. The solutions of the classical master
equation with suitable boundary conditions turn out to be generating
functionals for the gauge structure of the theory. $S$ is also the
starting point for the quantization.

Finally, the action $S\left[\Psi^{A},\Psi_{B}^{\star}\right]$ can
be expanded in a series in the antifields, while maintaining vanishing
ghost number and even Grassmann parity

\begin{eqnarray*} \label{s_bv}
	S_{BV}=S\left[\Psi^{A},\Psi_{B}^{\star}\right] & = & S_{0}+\phi_{i}^{\star}R_{\alpha}^{i}C^{\alpha}+C_{\alpha}^{\star}\frac{1}{2}T_{\beta\gamma}^{\alpha}\left(-1\right)^{\epsilon_{\beta}}C^{\gamma}C^{\beta}\\
	& + & \phi_{i}^{\star}\phi_{j}^{\star}\left(-1\right)^{\epsilon_{i}}\frac{1}{4}E_{\alpha\beta}^{ji}\left(-1\right)^{\epsilon_{\alpha}}C^{\beta}C^{\alpha}.
\end{eqnarray*}

When this is inserted into the classical master equation, one finds
that this equation implies the gauge structure of the classical theory.
In fact, this form is not unique but is the brief one for $S_{BV}$. One can turn back to the classical action $S_0$ when the antifields go to zero
\be
S_{BV}[\Psi, \Psi^\star]|_{_{\Psi^\star=0}} = S_0[\phi].
\ee

\subsection{Gauge Fixing and Quantization}

Although ghost fields have been incorporated into the theory, the solutions of classical master equation (\ref{eq:master}) have a set of invariances
\be
\frac{\p S}{\p z^a}R^a_b=0,
\ee 
with
\be
R^a_b=\omega^{ac}\frac{\p_l\p_r S}{\p z^c \p z^b}.
\ee
Due to these gauge freedoms the action (\ref{s_bv}), as a solution of classical master equation is not suitable for quantization via path integral and a gauge-fixing procedure is needed. The theory also contains many antifields that usually one wants to eliminate before computing amplitudes and S-matrix elements. One cannot simply set the antifields to zero because the action would reduce to the original classical action $S_0$ , which is not appropriate for starting perturbation theory due to gauge invariances. In the Batalin-Vilkovisky approach the gauge is fixed using a fermionic function which has Grassmann parity $\epsilon(\Theta)$=1, $\textbf{gh}[\Theta]=-1$ and is functional of fields $\Psi^A$ only. The antifields are eliminated through relation 
\be \label{fixing}
\Psi^\star_A=\frac{\p \Theta}{\p \Psi^A}
\ee
After implementing this gauge-fixing procedure we can define a surface in the functional space
\be
\Sigma_\Theta=\left\{\(\Psi^A, \Psi^\star_A\)|\Psi^\star_A=\frac{\p \Theta}{\p \Psi^A}\right\}.
\ee  
Hence for any functional $X\[\Phi,\Phi^\star\]$ we have
\be
X|_{_{\Sigma_\Theta}}=X\[\Psi,\frac{\p \Psi}{\p \Phi}\]
\ee
To construct a gauge-fixing fermion $\Theta$ of ghost number -1, one must again introduce additional auxiliary fields. The simplest choice utilizes a trivial pair $\bar{C}^\alpha$ and $\bar{\pi}^\alpha$ with the following properties
\bn
\epsilon\(\bar{C}^\alpha\)&=&\epsilon_\alpha+1, \quad \epsilon\(\bar{\pi}^\alpha\)=\epsilon_\alpha \no \\
\textbf{gh}\[\bar{C}^\alpha\] &=& -1 , \qquad \textbf{gh}\[\bar{\pi}^\alpha\]=0.
\en
The auxiliary fields $\bar{C}_\alpha$ are the Faddeev-Popov antighosts ($\bar{\pi}^\alpha$ are called Nakanishi-Lautrup fields)\footnote{Do not confuse antighost with anti-ghost.}. Along with these fields we include the corresponding antifields $\bar{C}^\star_\alpha$ and $\bar{\pi}^\star_\alpha$. Adding the term $\bar{\pi}^\alpha \bar{C}^\star_\alpha$ to the action $S$ does not spoil its properties as a proper solution to the classical master equation, and one obtains the \textit{non-minimal} action 
\be
S_{nm}=S+\bar{\pi}^\alpha \bar{C}^\star_\alpha.
\ee
We can think of these new auxiliary fields as a kind of Lagrange multipliers for the gauge-fixing terms. The simplest possibility for fermionic function $\Theta$ is
\be
\Theta=\bar{C}^\alpha \chi_\alpha\(\phi\)
\ee
where $\chi_\alpha$ are the gauge-fixing conditions for the fields $\phi$. The gauge-fixed action is denoted by
\be
S_\Theta=S_{BV-nm}|_{_{\Sigma_\Theta}}.
\ee
The quantum generating functional is defined by using the constraint (\ref{fixing}) to calculate the correlation function $X$ as
\be
I|_{_\Theta}\(X\)=\int \mathcal{D}\Psi \mathcal{D}\Psi^\star\delta\(\Psi^\star_A-\frac{\p \Theta}{\p \Psi^A}\)e^{\frac{i}{\hbar}W\[\Psi,\Psi^\star\]}X\[\Psi,\Psi^\star\].
\ee
Here $W$ is the quantum action, which reduces to S in the limit $\hbar\rightarrow 0$. An admissible $\Theta$ leads to well-defined propagators when the path integral is expressed as a perturbation series expansion. For a detailed discussion of the $W$ we refer the interested reader to the references \cite{mhct,Gomis:1994he}.  

\section {Acknowledgments} 

\ni V.N. would like to thank Prof. J. A. Helay\"el-Neto for valuable and insightful discussions. E.M.C.A. thanks CNPq (Conselho Nacional de Desenvolvimento Cient\' ifico e Tecnol\'ogico) through Grants No. 301030/2012-0 and No. 442369/2014-0, for partial financial support, CNPq is  a Brazilian scientific research support federal agency, and the hospitality of Theoretical Physics Department at Federal University of Rio de Janeiro (UFRJ), where part of this work was carried out. 

%\bibliographystyle{unsrt}
%\bibliography{Ref}

\end{document}